\documentstyle[eqsecnum,aps]{revtex}
\newcommand{\be}{\begin{equation}}
\newcommand{\en}{\end{equation}}
\newcommand{\bea}{\begin{eqnarray}}
\newcommand{\ena}{\end{eqnarray}}
\newcommand{\e}{\mbox{e}}
\newcommand{\tr}{\mbox{Tr}}
\newcommand{\trp}{\mbox{Tr}^+}
\newcommand{\trh}{\mbox{Tr}_{{\cal H}^l}}
\begin{document}
\begin{flushright}
\end{flushright}
\begin{center}
\large{\bf Fuzzy Actions and their Continuum Limits}
\bigskip

A.P. Balachandran$^{*}$, X. Martin$^{\dag}$ and Denjoe O'Connor$^{\dag}$ \\
\bigskip

${}^{*}$Physics Department, Syracuse University, \\
Syracuse, N.Y., 13244-1130, U.S.A.\\
${}^{\dag}$ Depto. de F\'{\i}sica, Cinvestav, \\
Apartado Postal 70-543,
M\'exico D.F. 07300, M\'exico.\\
\bigskip

\end{center}
\begin{abstract}
Previously matrix model actions for ``fuzzy'' fields have been
proposed using non-commutative geometry. They retained ``topological''
properties extremely well, being capable of describing instantons,
$\theta$--states, the chiral anomaly, and even chiral fermions with no
``doubling''. Here, we demonstrate that the standard scalar and
spinor actions on a $d$--dimensional manifold are recovered from such
actions in the limit of large matrices if their normalizations are
correctly scaled as the limit is taken.
\end{abstract}

\section{Introduction}
In contrast to conventional lattice discretizations, fuzzy physics
\cite{Connes,Madore,Landi,Varilly,Coquereaux,MSSVcommsphere} regulates
quantum fields on a manifold ${\cal M}$ by quantizing the latter,
treating it as a phase space. This method works because quantization
implies a short distance cut--off: the number of states in a volume
$V$ on ${\cal M}$ is infinite in classical physics, but becomes about
$V/\tilde\hbar^d$ on quantization, if $\tilde\hbar$ is the substitute
for Planck's constant and ${\cal M}$ has dimension $d$. The classical
limit $\tilde\hbar\rightarrow 0$ is then the continuum limit of
interest.

In past research,
\cite{Watamura1,Watamura2,GroPre,grklpr1,grklpr2,grklpr3,BalMonopolesSolitons,BalInstantons,BalFermionDoubling,FrGrRe}
several authors have explored this novel approach to discrete physics
and established that it can correctly reproduce continuum topological
features like instantons, $\theta$--states and the axial anomaly. Even
chiral fermions can be formulated without duplication \cite{BalInstantons}. When ${\cal M}$
is the two sphere $S^2$, this formulation shares common features with
the Ginsparg-Wilson approach \cite{BalFermionDoubling}.

Proposals have also been made in previous work for certain novel fuzzy
actions. They fulfill instanton-like bounds when such exist in the
continuum and in all cases are compatible with the known scaling
properties of the latter. They were conjectured to have correct
continuum limits as well. This paper verifies those conjectures for
the scalar and spinor fields. Gauge theories will be examined
elsewhere, but the correctness of the conjectures in all instances
looks plausible. The present work in this manner goes toward
establishing that the actions of
\cite{BalMonopolesSolitons,BalInstantons} in addition to retaining
important topological features, also have the correct continuum
limits. There is thus good reason to explore them further as
alternatives to conventional lattice models.

In section \ref{review}, we review the motivations and structure of
the proposed actions and indicate which of them we will study in this
paper. Introductory developments and useful asymptotic estimates
involving heat kernels methods are covered in sections
\ref{estimates}, \ref{elementary} and the Appendix. The remaining two
sections successfully employ this material to establish the continuum
limits. The concluding remarks are in section \ref{conclusion}.

\section{A review} \label{review}

Fuzzy physics is based on non-commutative geometry. In the continuum
limit $\tilde{\hbar}\rightarrow 0$, it is thus appropriate to compare
its action with those of conventional continuum physics written in the
language of non-commutative geometry. The proper fuzzy actions were in
fact inferred by their resemblance to the latter.

A central role is played in non-commutative geometry by the Dirac
operator. Let ${\cal D}$ be this operator for a manifold ${\cal M}$ of
dimension $d$. For a scalar field $\phi$ on ${\cal M}$, the free
action according to Connes is \be
S(\phi)=\trp \left( \frac{1}{|{\cal D}|^d} [{\cal D},\phi]^\dag
[{\cal D},\phi] \right) ,\label{scalara} \en
where $\phi$ and ${\cal D}$ are regarded as operators on the Hilbert
space of spinors and $\trp$ is the Dixmier trace \cite{VarGra}. It can
be explained as follows. Let $A$ be the operator within $\trp$ in
(\ref{scalara}) and $E^2$ be the eigenvalues of ${\cal D}^\dag {\cal
D}\equiv |{\cal D}|^2$. If $\tr_{|E|} (A)$ is the trace of $A$ up to
eigenvalues $|E|$ of $|{\cal D}|$, then \be
\trp (A)=\lim_{|E|\rightarrow \infty}\left( \frac{1}{\ln |E|}
\tr_{|E|} (A) \right) .\en
Connes shows that \be
S(\phi)= \int_{\cal M} |\nabla \phi|^2 d^d{\bf n},\en
where $\nabla$ is the covariant derivative operator in ${\cal D}$
and $d^d{\bf n}$ is the volume form on ${\cal M}$. In case ${\cal D}$
has zero modes, we will simply exclude the corresponding eigenspace
from all traces. 

There is a similar formulation of actions $S(\psi)$ for free spinors:
\be S(\psi)=\trp \left( \frac{1}{|{\cal D}|^d} \psi^\dag {\cal D}
\psi \right) .\label{spinora} \en
Actions appropriate for gauge theories will not be examined in this
paper. 

Under the scaling ${\cal D}\rightarrow \lambda {\cal D}$, the response
of these actions is \be
S(\phi)\rightarrow \lambda^{2-d} S(\phi), \ S(\psi)\rightarrow
\lambda^{1-d} S(\psi). \en
The critical dimensions for $\phi$ and $\psi$ to have scale invariant
actions are thus $d=2$ and $d=1$ respectively. The latter will not be
of interest to us since a line or a circle is not symplectic and
therefore cannot be quantized. 

There is an alternative form of great interest for these actions. Let
\be
\varepsilon=\mbox{sign}({\cal D}) \equiv \frac{\cal D}{|{\cal D}|} \en
be the sign of the Dirac operator. Then, as we will establish in subsequent
sections, \bea
S(\phi) & = & \frac{d-1}{d} \trp \left( \frac{1}{|{\cal D}|^{d-2}}
[\varepsilon,\phi]^\dag [\varepsilon,\phi] \right) ,\label{cscalara}
\\ S(\psi) & = & \frac{d-1}{d} \trp \left( \frac{1}{|{\cal D}|^{d-1}}
\psi^\dag \varepsilon \psi \right) .\label{cspinora}\ena

In fuzzy physics too, there is a Dirac operator with a proper
continuum limit. It is therefore natural to model its action on
(\ref{cscalara},\ref{cspinora}) or on
(\ref{scalara},\ref{spinora}). But as the fuzzy action involves only
finite dimensional Hilbert spaces, $\trp$ must be substituted by an
approximately normalised version of the normal trace. In that case,
the fuzzy versions of (\ref{cscalara},\ref{cspinora}) and
(\ref{scalara},\ref{spinora}) are no longer the same. There are
powerful topological reasons for preferring
(\ref{cscalara},\ref{cspinora}) over (\ref{scalara},\ref{spinora}) as
models for fuzzy versions. As has been shown elsewhere, only these
versions naturally fulfill instanton--like bounds and lead to
representations of $\theta$--states. Our task is thus to establish
that fuzzy models based on (\ref{cscalara},\ref{cspinora}) have the
correct limits.

By way of orientation, we now give a brief account of the fuzzy sphere
and its Dirac operator. Our discussion thereafter is a great deal more
general, but it is a useful illustration to keep in mind.

\subsection{The fuzzy sphere}
The sphere $S^2$ is a submanifold of $\Re^3$ given by \be
S^2=\{ {\bf n}\in \Re^3 : \sum_{i=1}^3 n_i^2 =1\} .\en
If $\hat{n}_i$ are the coordinate functions on $S^2$,
$\hat{n}_i ({\bf n})=n_i$, then $\hat{n}_i$ commute and the algebra
${\cal A}$ of smooth functions they generate is commutative.

In contrast, the operators $x_i$ describing $S_F^2$ are
non-commutative, their commutators being given by \be
[ x_i,x_j]=\frac{i\epsilon_{ijk} x_k}{\sqrt{l(l+1)}},\ \sum_{i=1}^3
x_i^2 ={\bf 1},\ l\in \{ \frac{1}{2},1,\frac{3}{2},\ldots \}
.\label{fcoord} \en
The $x_i$ approach $\hat{n}_i$ as $l\rightarrow\infty$. If $L_i=x_i
\sqrt{l(l+1)}$, then, from (\ref{fcoord}), it is clear that the $L_i$
give the irreducible representation (IRR) of $SU(2)$ with angular
momentum $l$. $L_i$ or $x_i$ generate the algebra $A_l=M_{2l+1}$ of
$(2l+1)\times (2l+1)$ matrices. 

A scalar product on $A_l$ is $<\!\xi
,\eta\!>=\tr(\xi^\dag \eta)$. $A_l$ acts on this Hilbert space by
left-- and right--multiplications, giving rise to the left-- and
right-- regular representations $A_l^{L,R}$ of $A_l$. For each $a\in A_l$,
we thus have two operators $a^{L,R}\in A^{L,R}$ acting on $\xi \in A_l$
according to $a^L\xi =a\xi$ and $a^R\xi=\xi a$. Note that
$a^Lb^L=(ab)^L$ but that $a^Rb^R=(ba)^R$. We assume by convention that
elements of $A_l^L$ are to be identified with fields or functions on
$S^2$. This identification is coherent with the fact that fields or
functions on $S^2$ come from ${\cal A}$.

Of particular interest are the angular momentum operators. There are
two kinds of angular momenta $L_i^{L,R}$ for $S_F^2$, while the
orbital angular momentum operator, which should annihilate ${\bf 1}$ is
${\cal L}_i=L_i^L-L_i^R$. $\vec{\cal L}$ plays the role of the
continuum $-i({\bf r}\times \nabla)$. The ``position'' operators are
not proportional to ${\cal L}_i$, but are instead
$L_i^L/\sqrt{l(l+1)}$.

The Dirac operator on $S_F^2$ is \be
D=\sigma\cdot {\cal L}+{\bf 1}.\en
It acts on $A_l\otimes {\cal C}^2=\{ (\psi_1,\psi_2):\psi_i\in A_l\}$.

$D$ admits a chirality $\gamma$ with which it anticommutes only after
state vectors of maximum total angular momentum $j=2l+1/2$ have been
projected out. This projection has no effect on our analysis however
and will therefore be ignored in the following.

\subsection{Preliminaries towards the Continuum}
Let $M_l$ be the fuzzy space for a manifold ${\cal M}$ of
dimension $d$ and let $D$ be its Dirac operator. The sign of $D$ is
\be
\epsilon=\frac{D}{|D|},\en
where it is understood that $0$--modes of $D$ are projected out
of the Hilbert space. The scalar and spinor field actions modeled on
(\ref{cscalara}) and (\ref{cspinora}) are then \bea
S_l (\phi) & = & \frac{1}{g_l} \tr_{{\cal H}^l} \left( \frac{1}{|D|^{d
-2}} [\epsilon, \phi]^\dag [\epsilon,\phi] \right),\label{cdscalara}\\ 
S_l(\psi) & = & \frac{1}{g_l} \tr_{{\cal H}^l} \left( \frac{1}{|D|
^{d-1}} \psi^\dag \epsilon \psi \right) ,\label{cdspinora}\ena
where $g_l$ is a logarithmically divergent (for 
$l\rightarrow\infty$ ) normalization, choosen such that
\be
\lim_{l\rightarrow\infty} \frac{1}{g_l} \tr_{{\cal H}^l} \left( \frac{1}{|D|
^{d}}\right)= \hbox{Volume of}({\cal M})
\en
Note that the scalar action $S(\phi)$ is scale invariant in the critical 
dimension $d=2$, which includes the case of the fuzzy sphere described
above. 

Operators and traces are associated with a finite dimensional Hilbert
space ${\cal H}^l$. For $S_F^2$, the dimension of ${\cal H}^l$ is
$2(2l+1)^2$. With increasing $l$, we have a family of nested Hilbert
spaces $\ldots\subset {\cal H}^l\subset{\cal H}^{l+1}\subset\ldots$
with a limiting Hilbert space ${\cal H}^\infty$. Now, as $l\rightarrow
\infty$, $D$ goes over to the continuum Dirac operator when both are
restricted to any finite--dimensional subspace. We can thus think of
operators and traces in $S(\phi)$ and $S(\psi)$ as being associated
with ${\cal H}^\infty$. As for $\phi$ and $\psi$, we require them to
become smooth continuum fields as $l\rightarrow \infty$. The exact
behavior we need is better stated in a coherent state basis which
will be done below.

We will consider the following more general forms of actions which
include (\ref{cdscalara}) and (\ref{cdspinora}) as special cases: \bea
S(\phi_0,\phi_1,\ldots,\phi_n) & = & \frac{1}{g_l} \tr_{{\cal H}^l} \left
( \frac{1}{|D|^p} \phi_0 [\epsilon, \phi_1]\ldots
[\epsilon,\phi_n] \right) ,\label{exas} \\ 
S(\psi_1,\psi_2) & = & \frac{1}{g_l}\tr_{{\cal H}^l} \left
( \frac{1}{|D|^p} \psi_1^\dag \epsilon \psi_2 \right), \label{exapp} \ena
where $\phi_i$ and $\psi_i$ are respectively fuzzy scalar and
spinor fields.

Let $\Lambda(l)$ be a cut--off eigenvalue of $|{\cal D}|$ which goes to
$\infty$ with $l$. Then $1/|{\cal D}_l|$ is defined as $1/|{\cal D}|$
whose modes with eigenvalue smaller than $1/|\Lambda(l)|$ are
projected out or exponentially suppressed, and $\varepsilon_l={\cal
D}/|{\cal D}_l|$. We can then approximate the above actions by \bea
S_l(\phi_0,\phi_1,\ldots,\phi_n) & = & \frac{1}{g_l} \tr_{{\cal
H}^\infty} \left ( \frac{1}{|{\cal D}_l|^p} \phi_0 [\varepsilon_l,
\phi_1]\ldots [\varepsilon_l,\phi_n] \right) ,\label{gencdscalara}\\ 
S_l(\psi_1,\psi_2) & = & \frac{1}{g_l} \tr_{{\cal H}^\infty} \left
( \frac{1}{|{\cal D}_l|^p} \psi_1^\dag \varepsilon_l \psi_2 \right),
\label{gencdspinora}\ena
as argued in the Appendix. We focus on these actions in what follows
as they are much easier to study.

The fuzzy spaces under consideration will be assumed to admit coherent
state representations. That is the case for $S^2_F$, ${\cal
C}P^N_F$ and many others. Let $|{\bf n},l\!>$ denote the coherent
state basis for ${\cal H}^l$, where ${\bf n}\in {\cal M}$, with the
normalization and conventions \bea
<\! {\bf n},l|{\bf n}',l\!> & = & \delta^d ({\bf n},{\bf n}')+o(1) \\
\int_{\cal M} d^d{\bf n} |{\bf n},l\!><\! {\bf n},l| & = & {\bf 1} .\ena
Here $\delta^d$ is the $\delta$ distribution on ${\cal M}$, $d^d{\bf n}$ is
the volume form and ${\bf 1}$ the identity operator. Then the
assumptions made on the fields are \bea
<\! {\bf n},l|\phi_i|{\bf n}',l\!> & = & \phi_i ({\bf n}) \delta^d({\bf
n},{\bf n}')+o(1) \label{phic}\\
<\! {\bf n},l|(\psi_i)_r|{\bf n}',l\!> & = & (\psi_i)_r ({\bf n})
\delta^d({\bf n},{\bf n}')+o(1) \label{psic} \ena
as $l\rightarrow\infty$. On the right hand side, $\phi_i$ and $\psi_i$
denote fields on the continuum manifold ${\cal M}$, and $(\psi)_r$ is
the $r$-th component of the spinor $\psi$. The use of the same symbols
$\phi_i$ and $\psi_i$ on both sides of (\ref{phic}) and (\ref{psic})
is not correct as they are matrices on the left hand side, but we
retain this notation for convenience. 

With (\ref{phic}), one has to leading order \be
<\! {\bf n},l|[\epsilon,\phi_i]|{\bf n}',l\!>={\cal D} {\cal G}_l
({\bf n}, {\bf n}') \left(\phi_i({\bf n})-\phi_i({\bf n}')\right) +
o(1) \label{epscom} \en
where ${\cal G}_l$ is the kernel of $1/|{\cal D}_l|$, \be
{\cal G}_l ({\bf n},{\bf n}')=<\! {\bf n}|\frac{1}{|{\cal D}
_l|}|{\bf n}'\!>,\en
$|{\bf n}\!>=|{\bf n},\infty \!>$ being the state vectors localized
at ${\bf n}\in {\cal M}$. Thus, it appears that the calculation of the
limit $l\rightarrow \infty$ requires finding the behavior of ${\cal
G}_l$ for large $l$.

\section{Preliminary estimates} \label{estimates} 
The necessary asymptotic estimates are best approached by adapting
heat kernel methods. The relevant heat kernel is \be
{\cal K}({\bf n},{\bf n}',t)=<{\bf n}\vert \e^{-{\cal D}^2t}\vert {\bf
n}'> .\en
It fulfills the heat equation \be
\partial_t {\cal K}({\bf n},{\bf n}',t)+\int d^d{\bf n}'' <\!{\bf
n}|{\cal D}^2|{\bf n}''\!> {\cal K}({\bf n}'',{\bf n}',t)=0\en
with the initial condition \be
{\cal K}({\bf n},{\bf n}',0)=\delta^d({\bf n},{\bf n}').\en

The operator $1/|{\cal D}_l|$, or more generally $1/|{\cal D}_l|
^\theta$, can be expressed in terms of this kernel, by using \be 
\frac{1}{|{\cal D}_l|^\theta}=\frac{1}{\Gamma(\frac{\theta}{2})}
\int_0^\infty \frac{dt}{t}t^{\theta/2}\e^{-{\cal D}_l^2t}.
\label{integrep}\en

Eigenvalues of ${\cal D} ^2$ exceeding $\Lambda(l)^2$ are cut off or
exponentially suppressed in $1/{\cal D}^2_l$. So we can approximate
the right hand side in (\ref{integrep}) by replacing ${\cal D}_l^2$ by
${\cal D}^2$ and restricting integration in $t$ to $t\geq T(l)\approx
1/\Lambda(l)^2$.  We thus have that \bea
\frac{1}{|{\cal D}_l|^\theta} & \simeq & \frac{1}{\Gamma(\frac{
\theta}{2})}\int_{T(l)}^\infty \frac{dt}{t}t^{\theta/2}\e^{-{\cal
D}^2t} \label{x}\\ 
{\cal G}_l^\theta ({\bf n},{\bf n}') & := & <{\bf n}\vert
\frac{1}{|{\cal D}_l|^\theta} \vert {\bf n}'> \frac{1}{\Gamma(
\frac{\theta}{2})} \int_{T(l)}^\infty \frac{dt}{t}t^{\theta/2} {\cal
K}({\bf n},{\bf n}',t). \ena
Furthermore, if $T_0$ is a (small) fixed number larger than
$T(l)$, one can write \be
{\cal G}_l^\theta ({\bf n},{\bf n}')=\frac{1}{\Gamma(\frac{\theta}{
2})} \left( \int_{T(l)}^{T_0}+\int_{T_0}^\infty \right)
\frac{dt}{t}t^{\theta/2} {\cal K}({\bf n},{\bf n}',t). \en
When $l\rightarrow \infty$, ${\cal G}_l^\theta ({\bf n},{\bf n}')$ has
a singularity as ${\bf n}\rightarrow {\bf n}'$. It comes from the
short--time  behavior of the heat kernel. Our interest is in this
singularity. Thus, the second integral, independent of $l$, can be
discarded and we can set (by an abuse of notation) \be
{\cal G}_l^\theta ({\bf n},{\bf n}')=\frac{1}{\Gamma(\frac{\theta}{
2})} \int_{T(l)}^{T_0} \frac{dt}{t}t^{\theta/2} {\cal K}({\bf n},{\bf
n}',t). \en 

At short times, the heat kernel has the asymptotic expansion \cite{Gilkey} \be
{\cal K}({\bf n},{\bf n}',t)=\frac{\Delta ({\bf n},{\bf n}')}{(4\pi t)^{d/2}}
\e^{-\frac{\sigma ({\bf n},{\bf n}')}{4t}} \left[ {\mathcal{I}}+\sum_
{n=1}^N {\cal A}_n ({\bf n},{\bf n}')t^n \right] +{\cal O}(t^{N+1}),
\label{Kexp} \en
where ${\cal I}$ is the identity operator, and the functions $\Delta$,
$\sigma$ and $A_n$ are independent of $t$ and have the short distance
behavior \be 
\Delta ({\bf n},{\bf n}')\rightarrow 1,\ \sigma({\bf n},{\bf n}')
\rightarrow |{\bf  n}-{\bf n}'|^2 \mbox{ and } {\cal A}_n ({\bf
n},{\bf n}') \rightarrow {\cal B}_{2n} ({\bf n}),\label{asymp} \en
when ${\bf n}\rightarrow {\bf n}'$. $B_{2n}$ are the Seeley
coefficients for the Dirac operator and $\sigma({\bf n},{\bf n}')$ is
the geodesic distance from ${\bf n}$ to ${\bf n}'$.

The corresponding asymptotic expansion of ${\cal G}_l^\theta$ is thus
\be 
{\cal G}_l^\theta ({\bf n},{\bf n}')=\frac{\Delta ({\bf n},{\bf n}
')}{(4\pi)^{d/2}\Gamma (\frac{\theta}{2})} \int_{T(l)}^{T_0}
\frac{dt}{t} t^{\frac{\theta-d}{2}} \e^{-{\sigma ({\bf n},{\bf n}')
\over 4t}} \left[ {\mathcal{I}}+\sum_{n=1}^N {\cal A}_n ({\bf n},{\bf 
n}') t^n \right] +{\cal O}(t^{N+1}). 
\label{integralG}
\en 
The leading behavior of this expression as $l\rightarrow \infty$, or
equivalently $T(l)\rightarrow 0$, and $|{\bf n}-{\bf n}'|\rightarrow 0$
can be found by letting $T_0$ also go to $\infty$: this only adds terms
well behaved as ${\bf n}\rightarrow {\bf n}'$. We find 
\bea
{\cal G}_l^\theta ({\bf n},{\bf n}')&=&\frac{2^{d-\theta}\Delta ({\bf
n},{\bf n}')}{(4\pi)^{d/2}\Gamma (\frac{\theta}{2})} \left[
{\Gamma(\frac{d-\theta}{2}) {\mathcal{I}}
\over(\sigma ({\bf n},{\bf n}'))^{\frac{d-\theta}{2}}}+\sum_{n=1}^N
{\Gamma(\frac{d-\theta-2n}{2}) {\cal A}_n ({\bf n},{\bf n}')\over
2^{2n}(\sigma ({\bf n},{\bf n}'))^{\frac{d-\theta-2n}{2}}} \right] .
\label{leadingasym}
\ena

For the fuzzy sphere $S_F^2$, this gives for the dominant term of
${\cal G}_l({\bf n},{\bf n}')$, i.e. ${\cal G}^1_l({\bf n},{\bf n}')$,
the expression \be
{\cal G}_l ({\bf n},{\bf n}')\simeq \frac{1}{2\pi}\frac{\mathcal{
I}}{|{\bf n}-{\bf n}'|} \en
so that the singular part of \be
\varepsilon ({\bf n},{\bf n}') := <\!{\bf n}|\varepsilon|{\bf n}'\!>=
\frac{1}{2\pi} {\cal D}_{\bf n} \left( \frac{\mathcal{I}}{|{\bf
n}-{\bf n}'|}\right) ,\en 
where the differentiation ${\cal D}_{\bf n}$ acts on the variable
${\bf n}$.

In the same way, for $d=4$, the singular part is \be
\varepsilon ({\bf n},{\bf n}')=\frac{1}{4\pi^2} {\cal D}_{\bf n}
\left( \frac{\mathcal{I}}{|{\bf n}-{\bf n}'|^3} + \frac{B_2({\bf
n})}{2|{\bf n}-{\bf n}'|} \right) .\en 
We will show that the second term, although divergent, does not
contribute to the continuum limit. In $d$--dimensions, the
corresponding expression is 
\be
\varepsilon ({\bf n},{\bf n}')=\frac{1}{(2\pi)^{d/2}} \sum_{n=0}
^{d/2-1} \frac{(d-3-2n)!! B_{2n}}{2^n|{\bf n}-{\bf n}'|^{d-1-2n}}, 
\label{vareps} \en
where $B_0=1$, $(-1)!!=1$, and $d$ must be even. If $d$ is odd, the
factors and $\Gamma (\frac{d-\theta-2n}{2})$ can have a pole for
$\theta=1$. This indicates that the cutoffs $T(l)\rightarrow 0$ 
an $T_0\rightarrow 0$ should be treated with more care. The terms with 
$n < (d-1)/2$ are not problematic and yield the leading singularities, 
while the term $n=(d-1)/2$ does not in fact yield a 
singularity as ${\bf n}\rightarrow {\bf n}'$.
Hence for odd $d$ we have 
\be
\varepsilon ({\bf n},{\bf n}')=\frac{1}{2\pi^{d/2}} \sum_{n=0}
^{{d-1\over2}-1} \frac{({d-1\over2}-n+1)! B_{2n}}{2^{2n}|{\bf n}-{\bf n}'|^{d-1-2n}}, 
\label{varepsodd} \en

\section{Elementary quantities on fuzzy spaces} \label{elementary}
Here we examine the asymptotics of certain basic expressions on the
fuzzy sphere.

Consider first \be
v_2=\mbox{Tr}_{{\cal H}^l}({1\over|D|^2}) .\en
The eigenvalues of $D$ are $\pm\lambda_j=\pm(j+1/2)$ for $j<2l+1/2$,
and $\lambda_{2l+1/2}=+(2l+1)$ for $j=2l+1/2$, each with
multiplicity $(2j+1)$. Thus \be
v_2=2\sum_{j=1/2}^{2l+1/2}{2j+1\over(j+1/2)^2}-\frac{2}{2l+1}
=4\psi(2l+1)-4\psi(1)-\frac{2}{2l+1}\en
where $\psi(x)=d\ln\Gamma(x)/dx$. For large $l$, 
$$v_2=4\ln(2l+1)+4\gamma_E-{4\over(2l+1)}+\dots ,$$
with $\gamma_E$ the Euler constant. Thus \be
{\displaystyle\lim_{l\rightarrow\infty}}{\pi\over\ln{l}} v_2=4\pi=
\mbox{Volume of } (S^2).\en

A slight generalization of this quantity in $d$ dimensions is \be
w(\phi)=\frac{(4\pi)^{d/2}\Gamma (\frac{d}{2})}{2 d_{\gamma}\ln(\Lambda
(l))} \tr_{{\cal H}^l}\left( \frac{1}{|D|^d}\phi \right) \en
where $\phi$ behaves like a smooth function on the manifold for large
$l$. For large $l$ then, using the Appendix, \be
w(\phi)\sim \frac{(4\pi)^{d/2}\Gamma (\frac{d}{2})}{2 d_{\gamma}
\ln(\Lambda (l))} \int_{\cal M} d^d{\bf n} \ \mbox{tr} [{\cal
G}_l^\theta ({\bf n},{\bf n}') ] \phi ({\bf n})\en
where $\mbox{tr}$ with lower case `t' indicates a trace over the
$\gamma$--matrices and $d_{\gamma}$ is the dimension of the
$\gamma$--matrices in $D$. This is \be
w(\phi)\sim \int_V d^d{\bf n} \phi ({\bf n}).\en
This is an important result and indicates how a potential
(typically a polynomial in fields) added to our free
actions will have the usual continuum limit. It also gives the asymptotic 
form 
\be
{1\over g_l}=\frac{(4\pi)^{d/2}\Gamma (\frac{d}{2})}{2 d_{\gamma}
\ln(\Lambda (l))} 
\en

\section{The scalar action}
We rewrite the scalar action (\ref{cdscalara}) as \be
S(\phi)=\frac{1}{g_l} \trh \left( \frac{1}{|D|^{d-2}} [\varepsilon,
\phi]^\dag [\varepsilon,\phi] \right) .\en
Its behavior for large $l$ can be deduced from that of $S_l(\phi_0,
\phi_1,\ldots,\phi_n)$ in (\ref{gencdscalara}). We consider the latter
for $n=3$, the critical dimension $d=2$, and the non--critical
dimension $d\not= 2$ separately. The treatment for a general $n$ is
similar but presents extra algebraic complexities.

\subsection{The critical dimension}
The behavior of $S(\phi)$ as $l\rightarrow \infty$ can be deduced
from that of $S_l$ in (\ref{gencdscalara}) for $p=0$ and $n=2$. For
large $l$, we get \be 
S_l(\phi_0,\phi_1,\phi_2)=\mbox{tr} \int d^2 {\bf n}_1\ d^2 {\bf n}_2 
\phi_0({\bf n}_1) <\! {\bf n}_1|[\varepsilon_l,\phi_1]|{\bf n}_2\!>
<\! {\bf n}_2|[\varepsilon_l,\phi_2]|{\bf n}_1 \!>.\en
Using Eq. (\ref{epscom}) where the kernel ${\cal G}_l$ is given in 
(\ref{integralG}), one finds in the leading term in large $l$ \bea
S_l (\phi_{0},\phi_{1},\phi_{2}) & = & \int d^2{\bf n}_1\ d^2{\bf
n}_2 \phi_0({\bf n}_1) {\cal D}_{{\bf n}_1} {\cal G}_l ({\bf n}_1,{\bf
n}_2) \left\{ (\phi_1 ({\bf n}_1)-\phi_1 ({\bf n}_2)) \right\} \\ 
&&\qquad\qquad\times {\cal D}_{{\bf n}_2} {\cal G}_l ({\bf n}_2,{\bf
n}_1) \left\{ (\phi_2 ({\bf n}_2)-\phi_2 ({\bf n}_1)) \right\}
.\label{int2d} \ena
We are interested in the (logarithmically) divergent term in this
expression for $l\rightarrow\infty$ which comes from the coincidence
limit of ${\cal G}_l$. This suggests the change of variables \bea
{\bf \xi} & = & {\bf n}_1-{\bf n}_2 \\
\overline{{\bf x}} & = & \frac{{\bf n}_1+{\bf n}_2}{2} \label{changev} \ena
with the expansions \bea
\phi ({\bf n}_1) & = & \phi (\overline{{\bf x}}+ \frac{{\bf
\xi}}{2})=\phi (\overline{{\bf x}})+\frac{{\bf \xi}^i}{2} \partial_i
\phi (\overline{{\bf x}})+{\cal O}_s({\bf \xi}^2) \label{nexp} \\
\phi ({\bf n}_2) & = & \phi (\overline{{\bf x}}-\frac{{\bf
\xi}}{2})=\phi (\overline{{\bf x}})-\frac{{\bf \xi}^i}{2} \partial_i
\phi (\overline{{\bf x}})+{\cal O}_s({\bf \xi}^2) \label{n1exp}\\
\phi ({\bf n}_1)-\phi ({\bf n}_2) & = & \xi^i \partial_i \phi
(\overline{{\bf x}})+{\cal O}_s ({\bf\xi}^2),\label{expansion} \ena
valid for any field $\phi$. Hence \be
{\cal D}_{{\bf n}_1}{ \cal G}_l ({\bf n}_1,{\bf n}_2)\left\{ (\phi_1 ({\bf
n}_1)-\phi_1 ({\bf n}_2)) \right\} = \left\{({\cal D}_{\bf \xi}{\cal
G}_l(|\xi|))\xi^i\right\} \partial_i \phi (\overline{{\bf x}})+{\cal
O}(|{\bf \xi}|^{1}). \label{asymcommepsphi} \en 

Putting all this together and using polar coordinates, for which
$d^2\xi=|\xi|d|\xi|d\Omega$, we find for $l$ large \be
S_l (\phi_{0},\phi_{1},\phi_{2})=N_l^{ij}\int d^2\overline{\bf x}
\left[ \phi_0 (\overline{{\bf x}}) \partial_i \phi_1 (\overline{{\bf x}})
\partial_j \phi_2 (\overline{\bf x})+{\cal O}_{ij} (|{\bf \xi}|)
\right] ,\en
with \be
N_l^{ij}=\mbox{tr}\int d|\xi||\xi|d\Omega \left\{{\cal D}_{\xi_k}{\cal
G}_l(|\xi|) {\cal D}_{\xi_k}{\cal G}_l (|\xi|)\right\} \xi^i\xi^j
. \en 
By rotational invariance, $N_l^{ij}=N_l\delta^{ij}$. Hence we get \be 
N_l=2\pi\int_0^{\infty}d|\xi||\xi|^3 \partial_{k}{\cal G}_l
(|\xi|)\partial^{k}{\cal G}_l (|\xi|). \en

The expansion (\ref{integralG}) shows that only the leading term
contributes to the logarithmic divergence. To isolate this divergence,
we use the leading term and find \be
N_l={2\over\pi^2}\int_{T(l)}^{T_0}dt_1\int_{T(l)}^{T_0}dt_2
{\sqrt{t_1t_2}\over(t_1+t_2)^3}.\en 
The indefinite form of this integral is given by polylogarithms \be
Li_{a}(z)={\displaystyle\sum_{k=1}^{\infty}} {z^k\over k^a} \en
in the form \be
\int dxdy{\sqrt{xy}\over(x+y)^3}={1\over2}{\sqrt{xy}\over(x+y)}
+{i\over4}\left\{Li_2(i\sqrt{{x\over y}})-Li_2(-i\sqrt{{x\over
y}})\right\} .\en
The asymptotic behavior of the polylogarithm for small $z$ is \be
Li_2(-i/z)=-1/2(\ln{z})^2-i{\pi\over2}\ln{z}-{\pi^2\over24}+\dots
\label{asympolylog}. \en
Thus \be
N_l={1\over2\pi}\ln l +{\cal O}(1). \label{divcrit} \en

Hence in the critical dimension $2$, \be
{\displaystyle\lim_{l\rightarrow\infty}} {\pi\over\ln(l)} S_l
(\phi_0,\phi_1,\phi_2)={1\over2}\int_{S^2}d^2{\bf n}\ \phi_0({\bf n})
\partial_i\phi_1({\bf n})\partial^i\phi_2 ({\bf n}). \en
It is the local scalar action once obvious choices of $\phi_i$ are
made. 

\subsection{The non--critical dimensions}
When $d>2$, there is no need to approximate $\varepsilon$ by
$\varepsilon_l$ as in (\ref{gencdscalara}): the Appendix shows that it
has the same leading large $l$ behavior as
(\ref{gencdscalara}). Instead, we can redefine $S_l$ to be \be 
S_l (\phi_0,\phi_1,\phi_2)=\mbox{Tr} ({1\over|D_l|^{d-2}}
\phi_0 [ \varepsilon,\phi_1 ] [ \varepsilon,\phi_2 ] ),\en
For large $l$ then, \bea
S_l(\phi_{0},\phi_{1},\phi_{2})=
\int d^d{\bf n}_1 \ d^d{\bf n}_2\ d^d{\bf n}_3
{\cal G}_l^{d-2}({\bf n}_1,{\bf n}_2)\phi_0({\bf n}_2) 
{\cal D}_{{\bf n}_2}{\cal G} ({\bf n}_2,{\bf n}_3)
\left\{(\phi_1 ({\bf n}_3)\right. \\ 
\qquad\qquad\left. -\phi_1 ({\bf n}_2))\right\}{\cal D}_{{\bf n}_3} {\cal
G} ({\bf n}_3,{\bf n}_1) \left\{ (\phi_2 ({\bf n}_1)-\phi_2 ({\bf
n}_3)) \right\} ,\ d>2.\label{int2} \ena
The kernels appearing here are divergent when $|{\bf n}_i-{\bf n}_j|
\rightarrow 0$ and give rise to the overall logarithmic
divergence. The following change of variable helps to isolate them: \bea
\xi_1 & = & {\bf n}_3-{\bf n}_2 \\
\xi_2 & = & {\bf n}_3-{\bf n}_1 \\
\overline{\bf x} & = & \frac{{\bf n}_1+{\bf n}_2+{\bf n}_3}{3}.\ena
We then find as $l\rightarrow \infty$, \be
S_l (\phi_{0},\phi_{1},\phi_{2})=N_l^{ij}\int d^d\overline{{\bf
x}} \phi_0 (\overline{{\bf x}}) \partial_i \phi_1 (\overline{
{\bf x}}) \partial_j \phi_2 (\overline{\bf x}) \label{inter1} \en
where, by spherical symmetry, \be
N_l^{ij}=N_l\delta^{ij}.\en
We find \bea
N_l & = & d_{\gamma} \int d{\bf \xi}_1 d{\bf \xi}_2 ({\bf \xi}_1 \cdot {\bf
\xi}_2) {\cal G}_a^{(d-2)}(|{\bf \xi}_1+{\bf \xi}_2|) \partial_i {\cal
G}^1(|{\bf \xi}_1|) \partial^i {\cal G} (|{\bf \xi}_2|)\\
& = &  \frac{4d_{\gamma}}{(4\pi)^{\frac{d+2}{2}}\Gamma (\frac{d-2}{2})}
\int_{1/T_0}^{1/T(l)} du \int_0^{+\infty} du_1 \int_0^{+\infty} du_2
\left( (1+u(u_1^2+u_2^2))^{-\frac{d+2}{2}} \right. \\
& & \left. +(d+2)u^2u_1^2u_2^2 (1+u(u_1^2+u_2^2))^{\frac{d+4}{2}}
\right) ,\label{inter2} \ena
where we made use of the leading terms of Eqs. (\ref{integralG}) and
(\ref{vareps}). The integrations over $u_1$ and $u_2$ can be done by
performing a double integration by parts on the second term and then
going to polar coordinates. The last integral, over $u$, isolates the
logarithmic divergence, \be
N_l \approx \frac{2d_{\gamma}\ln (l)}{(4\pi)^{d/2}\Gamma
(\frac{d}{2})} \ \frac{d-1}{d}.\en
It reduces to (\ref{divcrit}) in the limit of the critical dimension $d=2$.

Thus, for the general $d$, \be
{\displaystyle\lim_{l\rightarrow\infty}} \frac{(4\pi)^{d/2}\Gamma
(\frac{d}{2})}{2d_{\gamma}\ln(\Lambda)} \varphi_a (\phi_1,\phi_2,\phi_3)=
\frac{d-1}{d} \int_{V}d^d{\bf n} \phi_1({\bf n}) \partial_i\phi_2
({\bf n})\partial^i\phi({\bf n}) .\en  

\section{The spinor action}

As $d\geq 2$, we do not encounter the critical dimension $d=1$ of the
spinor action $S(\psi)$. It is enough for us to study
$S_l(\psi_1,\psi_2)$ for large $l$, as the limiting form of $S(\psi)$
can be deduced therefrom.

In the coherent state representation, $S_l$ reads \be
S_l (\psi_1,\psi_2)=\mbox{tr} \left( \int d^d{\bf n}_1 d^d{\bf n}_2
{\cal G}_l^{d-1}({\bf n}_1,{\bf n}_2) \psi_1^\dag ({\bf n}_2)
\varepsilon ({\bf n}_2,{\bf n}_1) \psi_2 ({\bf n}_1)\right) ,\en
where we have replaced $\varepsilon_l$ by $\varepsilon$, and the
spinors are assumed to be local in the continuum limit: \be
<\! {\bf n}|(\psi_i)_r|{\bf n}'\!>=(\psi_i)_r ({\bf n}) \delta ({\bf
n}-{\bf n}').\en 
It is easy to check the presence of a logarithmic divergence as $|{\bf
n}_1-{\bf n}_2|\rightarrow 0$. We can make it explicit by changing to
the variables $(\xi,\overline{\bf x})$ described in (\ref{changev}): \bea
S_l (\psi_1,\psi_2) & = & N_l^{ij} \left( \frac{\gamma_j}{2}
\int d\overline{\bf x} (\psi_1 (\overline{\bf x})\partial_i \psi_2
(\overline{\bf x}) -\partial_i \psi_1 (\overline{\bf x})\psi_2
(\overline{\bf x})) \right) ,\\
N_l^{ij} & = & -\delta^{ij} \frac{d_{\gamma}}{d} \int d^d{\bf \xi}\
{\cal G}_l^{d-1}(|{\bf \xi}|) {\bf \xi}^i \varepsilon_i (\xi).\ena 
Using the leading terms of Eqs. (\ref{integralG}) and (\ref{vareps}),
and going to spherical coordinates, we can evaluate $N_l^{ij}$ to get
\be N_l^{ij}=\delta^{ij} \frac{d-1}{d}\ \frac{2d_{\gamma}\ln (l)}{(4
\pi)^{d/2} \Gamma (\frac{d}{2})}\en
so that \be
{\displaystyle\lim_{l\rightarrow\infty}} \frac{(4\pi)^{d/2}\Gamma
(\frac{d}{2})}{2d_{\gamma}\ln(l)} S_l(\psi_1,\psi_2)=\frac{d-1}{d}
\int d^d{\bf n} \psi^\dag_1({\bf n}) {\cal D} \psi_2 ({\bf n}),\en 
which, up to the normalization, is the continuum Dirac action.

\section{Conclusion} \label{conclusion}
The fuzzy actions of \cite{BalMonopolesSolitons,BalInstantons} where
motivated by topological considerations. From their appearance, it is
not obvious that they have the desired continuum limits. In this
paper, we have established that these limits can be achieved for the
scalar and spinor actions if they are suitably scaled when the
cut--off $l$ approaches $\infty$. The scaling factors are
proportional to $(\ln (l))^{-1}$.

It remains to extend these considerations to actions of gauge theories
and to study the partition and correlation functions. The latter, at
least for the ``free field'' models considered here should be
accessible to analytic methods, these and the gauge field actions
should have the usual continuum limits. The study of the rate of
approach to these limits would be of particular interest.

\acknowledgements
This work was supported by the joint NSF-CONACyT grant number
$E120.0462/2000$. A.P. Balachandran and D. O'Connor were also
supported respectively by the DOE grant $DE-FG02-85ER40231$, and the
CONACyT grant $30422-E$. X. Martin was supported by CONACyT and
SNI-M\'exico. 

\section*{Appendix}
In this Appendix will be explained how the functionals on the
non-commutative spaces whose limits are to be studied can be turned
into integrals on the limiting commutative manifold. For that, it will
be convenient to suppose that each of the non--commutative algebras
$M_l$ admit a coherent states representation.

Coherent states form a complete set of vectors $|{\bf n},l\! >$ on the
algebra indexed by points ${\bf n}$ on the limiting manifold ${\cal
M}$. They have the asociated resolution of the identity \be
\int_{\cal M} d^d{\bf n} |{\bf n},l\! ><\!{\bf n},l| =1\en
and are orthogonal in the limit $l\rightarrow \infty$: \be
<\!{\bf n},l|{\bf n}',l\! >=\delta^d({\bf n}-{\bf n}')+o(1).\en
This coherent state representation naturally generates a linear
imbedding of the non-commutative algebras $M_l$ into the algebra of
functions on ${\cal M}$: \be
\varphi_l \in M_l \mapsto \varphi ({\bf n})=<\!{\bf n},l|\varphi_l|
{\bf n},l\! >.\en
In general, and that is certainly true in the case of a compact simple
Lie group, this mapping is injective. This is because the coherent
states form an overcomplete set, so that the diagonal terms $\varphi
({\bf n})$ are sufficient to reconstruct completely the initial matrix
$\varphi_l$.  In the following, we will assume that it {\it is}
possible to reconstruct any field from its diagonal elements in the
coherent state basis.

This mapping is not a mapping of algebras since it can not map a
non-commutative multiplication to a commutative one. The product
obtained on the algebra of functions by mapping the product on the
non-commutative algebras will be called $*$--product. It is thus
defined as \be 
<\!{\bf n},l|\varphi_l\psi_l|{\bf n},l\! >=\varphi ({\bf n})*\psi 
({\bf n}).\en
With the assumption above, the $*$--product goes over to the normal
product on the manifold in the commutative limit $l\rightarrow\infty$,
with corrections which go like $1/l$.  In a way, this property means
that fields become local in the continuum limit.

Something similar can also be derived with operators other than
multiplication. Indeed, if ${\cal O}_l$ is an operator on the Hilbert
space ${\cal H}_l$, it is mapped to a kernel \be
{\cal O} ({\bf n},{\bf n}')=<\!{\bf n},l|{\cal O}|{\bf n}',l 
\! > \en
on the manifold ${\cal M}$. In this case, there is no particular
reason to think that the diagonal elements of the operator should be
sufficient to describe it completely. Then composing this operator
with a left multiplication operator will map to another kind of
product which will be denoted as $\#$ \be 
<\!{\bf n}|{\cal O}\phi_l|{\bf n}'\! >={\cal O} ({\bf n},{\bf
n}')\# \phi ({\bf n}').\en 
It is a reasonable assumption, which will be made in the following,
that this product also goes over to the standard product in the
large $l$ limit, with corrections of order $1/l$.

At this point, taking for instance the functional (\ref{exapp}) with
$n=2$, one can write \bea
S_l(\psi_1,\psi_2) & = & \frac{1}{g_l} \trh \left( \frac{1}{|D|^p}
\psi_1\dag \epsilon \psi_2 \right) \\
& \sim & \frac{1}{g_l} \int_{\cal M} d^d{\bf n}_1 d^d{\bf n}_2 <\! {\bf
n}_1,l|\frac{1}{|D|^p}|{\bf n}_2,l\!>\# \psi^\dag_1 ({\bf n}_2)\# <\!
{\bf n}_2,l|\epsilon|{\bf n}_1,l\!>\# \psi_2  ({\bf n}_1) ,\label{exa}\ena
where the sequence $g_l$ is expected to diverge logarithmically to get
a finite result for the continuum action, and $\sim$ means that the
two expressions are equivalent when $l\rightarrow\infty$.

Because of the ansatz on the limit of the $\#$--product made above, it
should be clear that the expected logarithmic divergence of
(\ref{exa}) will not be affected if $\#$--products are replaced by
standard products. This yields immediately \be
S_l(\psi_1,\psi_2)\sim\frac{1}{g_l} \int_{\cal M} d^d{\bf n}_1 d^d{\bf n}_2
<\! {\bf n}_1,l|\frac{1}{|D|^p}|{\bf n}_2,l\!> \psi^\dag_1 ({\bf n}_2)
<\! {\bf n}_2,l|\epsilon|{\bf n}_1,l\!> \psi_2  ({\bf n}_1) .\en
At this point, what remains to be done is to evaluate the limiting
behaviour of the kernels for $1/|D|^p$ and $\epsilon$ which appear on
the right hand side.

It is reasonable to think that, having the same eigenvalues and
eigenfunctions with the same structure, the discrete operators
$1/|D|^p$ and $\epsilon$ should converge to their equivalent
counterparts $1/|{\cal D}_c|^p$ and $\varepsilon_c$ whose eigenvalues
larger than the cut--off eigenvalue $\Lambda(l)$ have been set to
zero. Thus, \be 
S_l(\psi_1,\psi_2)\sim\frac{1}{g_l} \int_{\cal M}d^d{\bf n}_1 d^d{\bf n}_2
<\!{\bf n}_1|\frac{1}{|{\cal D}_c|^p}|{\bf n}_2\!> \psi^\dag_1 ({\bf n}_2)
\varepsilon_c ({\bf n}_2,{\bf n}_1) \psi_2  ({\bf n}_1).\en

These truncated operators in the continuum are still difficult to
describe since their cut--off is spectral. However, in the limit
$l\rightarrow \infty$, these truncated operators converge weakly (i.e
their action on any given spinor converges to the action of the weak
limit) to the usual Dirac kernels $1/|{\cal D}|^p$ and
$\varepsilon$. This is simply because these operators are bounded and
therefore their high frequency behaviour is irrelevant. Indeed,
calling generically ${\cal O}$ and ${\cal O}_c$ the operator
considered and its truncation respectively, \be
({\cal O}-{\cal O}_c) (\sum_{i=0}^\infty c^j_{ms}E_{ms}^j)={\cal O}
(\sum_{i=\Lambda(l)}^\infty c^j_{ms}E_{ms}^j)\leq |||{\cal O}|||\ ||
\sum_{i=\Lambda(l)}^\infty c^j_{ms}E_{ms}^j||\rightarrow 0,\en
where $|||\cdot |||$ denotes the norm of the operator ${\cal O}$. Note
that this shows that operators of the form $[\varepsilon_c,\phi]$ or 
$\varepsilon_c\psi$ also converge weakly to the corresponding
untruncated operators. 

This weak convergence is however insufficient for our purpose because
we know that the expressions to be studied are logarithmically
divergent when $l\rightarrow\infty$ and therefore that the high
frequency behaviour of the operators {\it is} important. Still, as
long as one truncated operator is kept, the large $l$ behaviour of the
others will not matter. Keeping the truncated kernel $1/|{\cal D}_c|^p$, we
therefore have for the example considered, \be
S_l(\psi_1,\psi_2)=\frac{1}{g_l} \int_{\cal M} d^d{\bf n}_1 d^d{\bf n}_2
<\!{\bf n}_1|\frac{1}{|{\cal D}_c|^p}|{\bf n}_2\!> \psi^\dag_1 ({\bf n}_2)
\varepsilon ({\bf n}_2,{\bf n}_1) \psi_2  ({\bf n}_1).\label{exa2}\en

Now for the remaining operator $1/|{\cal D}_c|^p$, the high frequency
behaviour {\it does} matter and it will serve to regularise the
expressions we are interested in. A good approximation for $1/|{\cal
D}_c|^p$ is the operator $1/|{\cal D}_l|^p$, introduced in
(\ref{gencdscalara}) and defined in (\ref{integrep}), whose high
frequency eigenvalues are made to decrease exponentially. To compare
it to the truncated operators, their eigenvalues should be
compared. By construction, all these operators are functions of the
Dirac operator and therefore have the same eigenfunctions ${\cal
E}^j_{ms}({\bf n})$, and the eigenvalues $\Delta_j$ of their
difference are thus just the difference between their respective
eigenvalues. We find for the ``relative'' difference when $E_j\le
\Lambda (l)$: \bea
E_j^p\Delta_j & = & \frac{E^p_j}{\Gamma({\frac{p}{2}})}\int_0^
{T(l)} \frac{dt}{t}t^{p/2} \e^{-E_j^2t} \label{eigenlow}\\ 
& \le & \frac{1}{\Gamma(\frac{p}{2})} \int_0^{T(l)E_j^2} \frac{du}{u}
u^{p/2} \le \frac{1}{\Gamma(\frac{p+2}{2})} \left( \frac{E_j}{\Lambda
(l)}\right) ^p .\label{boundlow}\ena 
In the other case $E_j> \Lambda (l)$, the eigenvalues of $1/|{\cal
D}_c|$ are zero, and the eigenvalues $F_j$ of $1/|{\cal D}_l|$ are
given by \bea
F_j & = & \frac{1}{\Gamma({\frac{p}{2}})}\int_{T(l)}
^{+\infty}  \frac{dt}{t}t^{p/2} \e^{-E_j^2t} \\
& \leq & \frac{1}{E_j^p \Gamma(\frac{p}{2})} \e^{-T(l)E_j^2/2}
\int_0^{+\infty} \frac{du}{u}u^{p/2} \e^{u/2} \leq \frac{2^{p/2}}{E
_j^p}\e^{-(E_j/\Lambda (l))^2/2} .\label{boundhigh} \ena
So, the regularised kernel $1/|{\cal D}_l|^p$ is equivalent to the
truncated one whenever $E_j\ll\Lambda(l)$, and is exponentially
suppressed when $E_j\gg\Lambda(l)$. However, the difference between
the two kernels becomes of order one near the cut--off eigenvalue
$\Lambda(l)$. 

Introducing a complete set of Dirac eigenfunctions in the functional
(\ref{exa2}) and the one we want to replace it by, namely
(\ref{gencdspinora}), we get \bea
S_l(\psi_1,\psi_2) & \sim & \frac{1}{g_l} \sum_{j=0}^{\Lambda(l)}\frac{1}{E
_j^p}\int_{\cal M} d^d{\bf n}_1 d^d{\bf n}_2 {{\cal E}^j_{ms}}^\dag
({\bf n}_2) \psi^\dag_1 ({\bf n}_2) \varepsilon ({\bf n}_2,{\bf n}_1)
\psi_2 ({\bf n}_1) {\cal E}^j_{ms}({\bf n}_1),\label{acts} \\
\tr_{{\cal H}^\infty}\left( \frac{1}{|{\cal D}_l|^p}\psi_1^{\dag}
\varepsilon\psi_2 \right) & = & \frac{1}{g_l} \sum_{j=0}^{\infty}
\frac{1}{F_j} \int_{\cal M} d^d{\bf n}_1 d^d{\bf n}_2 {{\cal
E}^j_{ms}}^\dag ({\bf n}_2) \psi^\dag _1 ({\bf n}_2) \varepsilon ({\bf
n}_2,{\bf n}_1) \psi_2 ({\bf n}_1){\cal E}^j_{ms}({\bf n}_1).
\label{actl}\ena
For modes $j\leq \Lambda_-(l)\ll \Lambda(l)$, the inequalities
(\ref{boundlow}) suggest that the difference between these two
expressions should be subdominant. For modes $j\geq
\Lambda(l)$, the inequalities (\ref{boundhigh}) suggest that the
contribution in Eq. (\ref{actl}) should also be subdominant since all
modes are exponentially suppressed. Finally for intermediate modes
$\Lambda_-(l)\leq j\leq \Lambda(l)$, the relative contribution to the
spinor type action in Eq. (\ref{acts}) is of order
$\ln [\Lambda(l)/\Lambda_-(l)]/\ln(\Lambda(l))$ since we know that it
diverges logarithmically in the large $l$ limit. For an appropriate
choice of $\Lambda_-(l)$, this relative contribution can be made
subdominant. Thus, it is to be expected that the two functionals
(\ref{acts}) and (\ref{actl}) have the same continuum limit
$l\rightarrow\infty$. 

Hence, we have found that, with a sequence $g_l$ which diverges
logarithmically, \be
\lim_{l\rightarrow\infty} \left( S_l(\psi_1,\psi_2)\right) =\lim_{l
\rightarrow\infty} \left(\frac{1}{g_l} \int_{\cal M} d^d{\bf n}_1
d^d{\bf n}_2 {\cal G}_l^p ({\bf n}_1,{\bf n}_2) \psi^\dag_1 ({\bf
n}_2) \varepsilon ({\bf n}_2,{\bf n}_1) \psi_2 ({\bf n}_1) \right),\en 
which is exactly the form (\ref{gencdspinora}) we were looking for.
For the other type of functionals, the scalar type actions (\ref{exas}),
the same reasoning will yield that they are equal to the expression
(\ref{gencdscalara}) we studied in the article.

\bibliographystyle{unsrt}

\end{document}